\documentclass{epl}

\usepackage{epsf}

\newcommand{\szz}{\sigma_{zz}}
\newcommand{\sxx}{\sigma_{xx}}

\newcommand{\sxz}{\sigma_{xz}}

\newcommand{\osl}{\textsc{osl}}

\newcommand{\De}{\Delta}
\newcommand{\ra}{\right >}
\newcommand{\la}{\left <}

\begin{document}

\title{Stress response function of a two-dimensional ordered\\
       packing of frictional beads}
\shorttitle{Stress response function etc...}
\author{
Laurent Breton\inst{1}      \and   Philippe Claudin\inst{2}    \and
\'Eric Cl\'ement\inst{2} \\ \and   Jean-Daniel Zucker\inst{1}}
\shortauthor{L. Breton \etal}
\institute{
\inst{1}
Laboratoire d'Informatique de Paris 6 - p\^ole I.A. (UMR 7606),\\
4 place Jussieu - case 169, 75252 Paris Cedex 05, France.\\
\inst{2}
Laboratoire des Milieux D\'esordonn\'es et H\'et\'erog\`enes (UMR 7603),\\
4 place Jussieu - case 86, 75252 Paris Cedex 05, France.}


\maketitle

\begin{abstract}
We study the stress profile of an ordered two-dimensional packing of beads
in response to the application of a vertical overload localized at its top
surface. Disorder is introduced through the Coulombic friction between
the grains which gives some indeterminacy and allows the choice of one
constrained random number per grain in the calculation of the contact forces.
The so-called `multi-agent' technique we use, lets us deal with systems as
large as $1000\times1000$ grains. We show that the average response profile
has a double peaked structure. At large depth $z$, the position of these peaks
grows with $cz$, while their widths scales like $\sqrt{Dz}$. $c$ and $D$ are
analogous to `propagation' and `diffusion' coefficients. Their values depend
on that of the friction coefficient $\mu$. At small $\mu$, we get
$c_0-c \propto \mu$ and $D \propto \mu^\beta$, with $\beta \sim 2.5$, which
means that the peaks get closer and wider as the disorder gets larger. This
behavior is qualitatively what was predicted in a model where a stochastic
relation between the stress components is assumed.
\end{abstract}

The statics of granular materials is subject to an active research, see e.g.
\cite{PGDG}. One of the main issues concerns the link between the distribution
of stresses in a granular medium and its `past history' which has induced, at
the microscopic level, a certain texture to the packing. As a matter of fact,
the mechanical properties of an assembly of grains depends on the way it has
been prepared. A now famous example is that of the sandpile: when
built from the source point of a hopper, the pressure profile at the bottom
of a pile shows an unambiguous minimum right below its apex \cite{Smid}. By
contrast, this profile is almost flat with a slight hump when the pile has
been prepared by successive horizontal layers \cite{Vaneltas}. The difference
between the two sandpiles could be seen at the level of the grains whose
contacts and forces are oriented in relation to the external solicitation
applied: during the construction, grains tend to gain strong
contacts -- i.e. carrying a large force -- in the direction of compression
\cite{Radjai}. This strong texture can be visualized in photoelasticity
experiments which show clear `force chains' structures \cite{Dantu}. The
minimum of pressure at the bottom of a sandpile can then be interpreted
as a screening effect of the weight of the grains by these chains.

Conceptually, the simplest test to probe the internal packing structure of a
layer of grains is to perform a stress response function: the layer is
submitted to a given force $\vec{F}$ localized at a point of its top surface,
and the resulting additional stress is measured at some distance from that
point. $\vec{F}$ must be small and supportable by the packing, in order to
avoid contact rearrangements. The interesting point is that the shape of the
averaged response function gives some information on the grain packing.

Let us take few instructive examples. Recent experiments were performed
with layers of natural sand submitted to a vertical overload \cite{manip3D}.
The response pressure at the bottom of the layer presents a single centered
peaked profile whose width scales linearly with the layer thickness. Such
shape and scaling resemble that of the isotropic elastic response. However the
measurements have shown that these profiles are very dependent on the system
preparation: compacted sand layers have a much wider response peak than loose
ones \cite{repelas}. Other experiments were performed on 2d pilings of
photoelastic grains \cite{manip2D}, showing the importance of the
amount of disorder. The stress response function of a regular packing of
monodiperse beads has a double peaked structure localized on the two
diagonals of the triangular lattice -- the behavior in 3d ordered
systems is similar \cite{Chicago}. Polydispersity in the distribution of the
grain diameters make these two peaks move closer to each other. They
eventually merge into a single one at large disorder. Numerical results are
also available in 2d: the stress response function of a polydisperse assembly
of frictionless discs was computed using the generic isostatic property of
such systems \cite{Head}. The corresponding response profile has two peaks
that become asymmetrical when the packing is initially sheared. By contrast,
Contact Dynamics simulations of a disordered layer of frictional pentagons
show a single peaked response, although the distribution of contact
orientation has two clear oblique preferred directions \cite{Moreau}.

The stress response function can also be used as a test to discriminate
between all the models which aim at describing the statics of granular
materials. As a matter of fact, the different classes of models give
qualitatively different response profiles. Models with equations of the
elliptic class, like those from elastic theories, typically lead to single
peaked shapes, with peak widths scaling like the depth $h$. They can however
also give double peaked response profiles if some strong anisotropy is included
\cite{GGetCie}, but these profiles keep a linear scaling with $h$ \cite{Otto}.
In models which handle scalar variables like the $q$-model \cite{qmodel},
response profiles also show a single peak, but narrower: its width grows like
$\sqrt{h}$ only. Note that such a behavior was claimed to be observed,
but on a quite particular and rather small scale `brick' packing
\cite{daSilva}. Initially proposed to explain the central minimum of pressure
below a sandpile built from a hopper, the models `\osl' (for `Oriented Stress
Linearity') where one postulates, as a closure relation, a phenomenological
linear relationship between the stress tensor components \cite{bcc,wcc} of the
type $\sxx = \eta \szz + \eta' \sxz$, give by contrast hyperbolic equations.
Such equations are formally identical to wave propagative equations and have
so-called `characteristics lines' along which stresses are transmitted. The
coefficients $\eta$ and $\eta'$ encode here the texture of the system they
describe and thus depend on the way the packing was prepared -- they are
`history dependent'. As a consequence, the corresponding response functions
have two narrow peaks resulting from the two characteristics initiated at the
overload point. Edwards and Grinev who studied isostatic assemblies of grains
with infinite friction obtained a `stress-geometry' equation which, in the
simplest case, is equivalent to the \osl\ relation written above
\cite{Edwards}, see also \cite{Ball}. Tkachenko and Witten also deduced a
large scale relation of the \osl\ type from microscopic quantities, but this
time on isostatic packings of frictionless beads \cite{Tkachenko}. This is
consistent with the result of Head \etal\ cited above, but rather
contradictory to the conclusions of the extensive work of Roux on the
mechanical properties of isostatic systems \cite{Roux}.

\begin{figure}[t]
\begin{center}
\epsfxsize=0.37\linewidth
\epsfbox{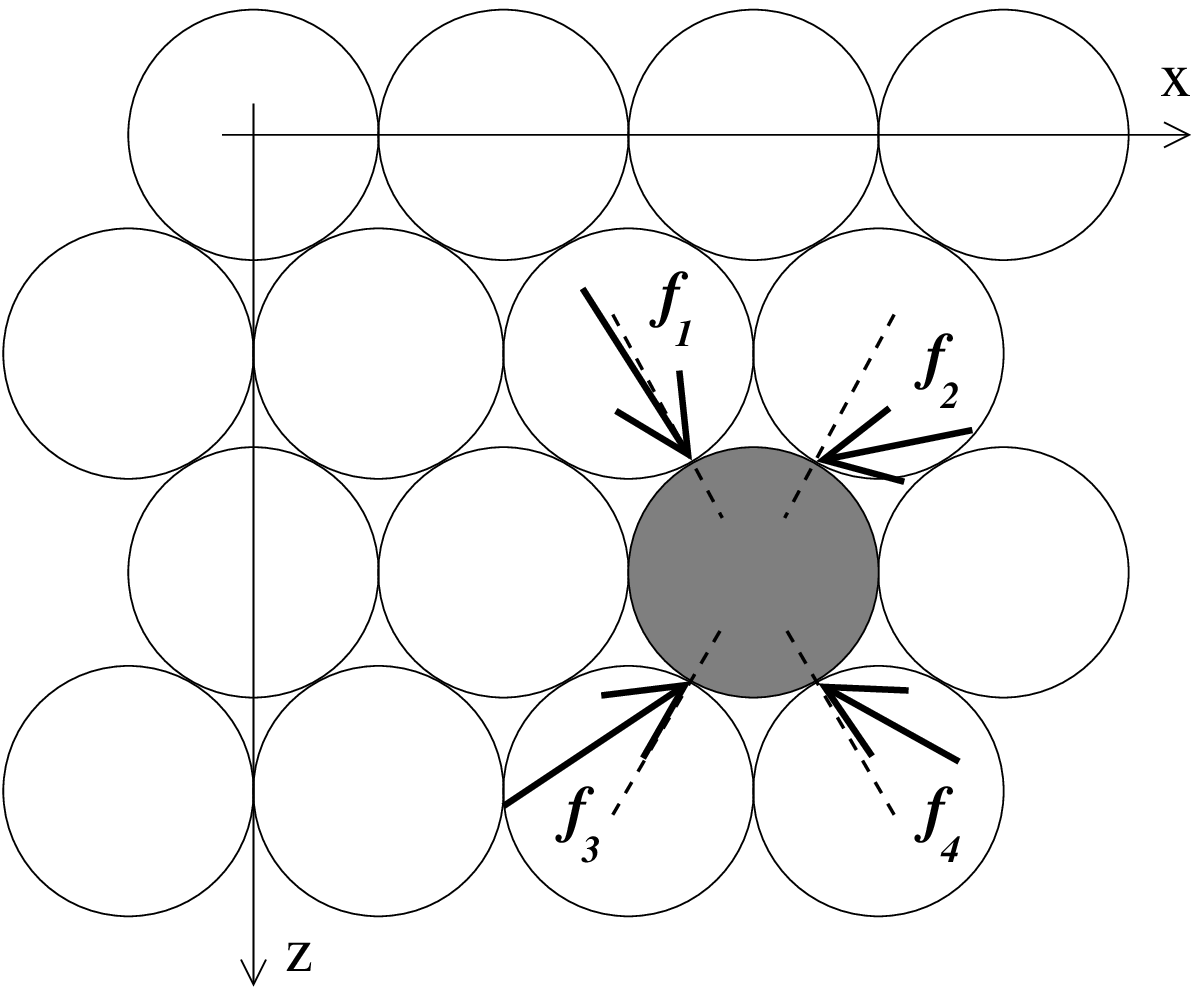}
\hspace*{1cm}
\epsfxsize=0.37\linewidth
\epsfbox{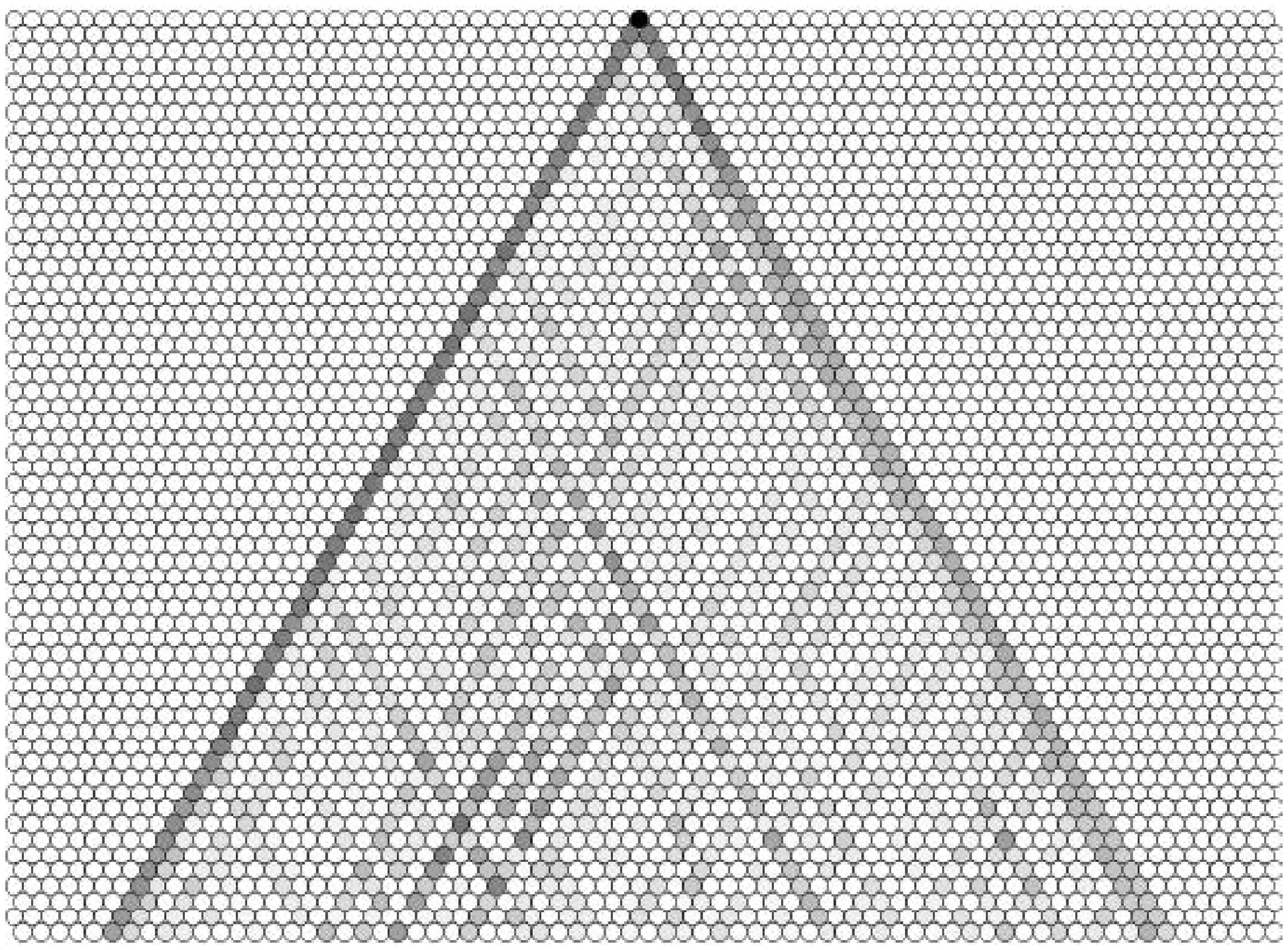}
\end{center}
\caption{
\underline{Left}: Grains are placed on a regular triangular lattice, but
have four contacts with their upper and lower neighbors only. $x$ is the
horizontal axis, and $z$ measures the depth from the surface. All lengths are
measured in units of grains diameters. Only compressive forces are admissible
for these cohesionless grains. The tangent of the angle $\theta_i$
that the force $\vec{f_i}$ makes with the contact direction (dashed lines)
cannot exceed the Coulomb friction coefficient $\mu$. These forces must
satisfy the equilibrium relations
$\vec{f_1} + \vec{f_2} + \vec{f_3} + \vec{f_4} = 0$ and
$f_1\sin\theta_1 + f_2\sin\theta_2 + f_3\sin\theta_3 + f_4\sin\theta_4 = 0$,
leaving one undetermined unknown per grain chosen at random.
\underline{Right}: Forces on a $70\times60$ piling due to a vertical
overload localized at its top surface. Forces are larger when the grains are
darker. One can clearly see the double peaked response. This picture has
been computed with $\mu=0.1$.
}
\label{reseau}
\end{figure}

In this context, Eloy and Cl\'ement have proposed few years ago to study the
statics of a regular two-dimensional layer of beads \cite{Eloy}. Although in
this system each grain has six neighbors, only upper and lower contacts were
considered -- see figure \ref{reseau} (left). For frictionless beads, all
contact forces $\vec{f_i}$ must be along the normal at the contact point, and
it is easy to see that they can all be uniquely computed layer by layer, from
the force balance equations and the top boundary conditions. More interesting
is the case where each bead contact is subject to a Coulombic friction
condition $|f_i^T| \le \mu f_i^N$, where $f_i^T$ and $f_i^N$ are respectively
the tangential and the normal components of the contact force $\vec{f_i}$.
$\mu=\tan\phi$ is the friction coefficient: all forces have to be in the
Coulomb cone of angle $\phi$. The force and torque balance equations give for
each grain three relations, while there are four unknowns -- note that each
contact belongs to two adjacent beads. The idea is to explore the space of
mechanically acceptable solutions by choosing one of the four unknowns
at random among the values permitted by the friction conditions. Besides, we
also impose that all contact forces have to be positive. Therefore, this
simple toy model, where disorder is governed by friction only, gives a clear
framework where one can compute contact forces.

The stochastic calculation begins at the surface $z=0$ where overload forces
are given, and continues deeper and deeper, one layer after another. Sometimes,
the interval of admissible values in which the random number must be chosen
is empty and thus makes the choice of this random number impossible. It means
that for given upper forces on such a grain, at least one of the lower forces
would be out of its Coulomb cone, or that a contact force is negative. A
rearrangement has then to occur. In the original work of \cite{Eloy}, the
random numbers of the $n$ previous layers were simply recomputed until the
calculation could go on, where $n$ was chosen proportionally to the number of
failures encountered in the current layer. Such a procedure is of course very
slow and actually makes the total computation time increase exponentially with
both the size of the system and the value of the friction. The pile could then
be entirely rebuilt hundreds of times before having the chance to generate no
impossibilities at all. In practice, it was not possible to built, in a
reasonable \textsc{cpu} time, piles larger than $50$ grains for $\mu \le 0.6$,
which makes statistical and large scale studies difficult. On these rather
small pilings, \osl\ features were evidenced \cite{Eloy}.

A much more sophisticated method called \textsc{GranuSolve} \cite{Breton}
involving a so called `multi-agent' representation and eco-solving algorithms
\cite{drogoul} have been used here to improve the computation of the model.
In this method, we consider each bead as an `agent' whose `goal' is to reach
its static equilibrium with respect to its mechanical conditions, i.e. to
`solve itself'. When a grain is solved, it communicates its contact forces to
its lower neighbors. The lower bead-agents continue the process, and solve
themselves until a failure is encountered by one of them. In such a case, the
bead-agent asks its upper neighbors to change theirs own values for the
contact forces they have in common. The rearrangements are therefore treated
locally by the grains and we do not need to recompute all random numbers from
the upper layers but only in the local area from where the failure occurred.
When averaging the data, all configurations are taken with the same weight.
To avoid any bias in the scan of the space of solutions,
we were particulary carreful in the choice that the simulation makes for the
next bead-agent that needs to get priority treatment. The choice that we
finally kept was to solve grain layers from top to bottom, starting on a new
random grain on each layer, and treating unsolved grains as `clusters'. With
this technique, the computation time is linear with the number of grains,
and a complete resolution of a $1000\times1000$ pile takes few minutes on a
G4 400 MHz machine, up to friction coefficients as large as $\mu\sim 1.7$. We
shall however restrict our discussion to relatively small values of $\mu$
for which averaged data confidence is high.

\begin{figure}[t]
\begin{center}
\epsfxsize=0.43\linewidth
\epsfbox{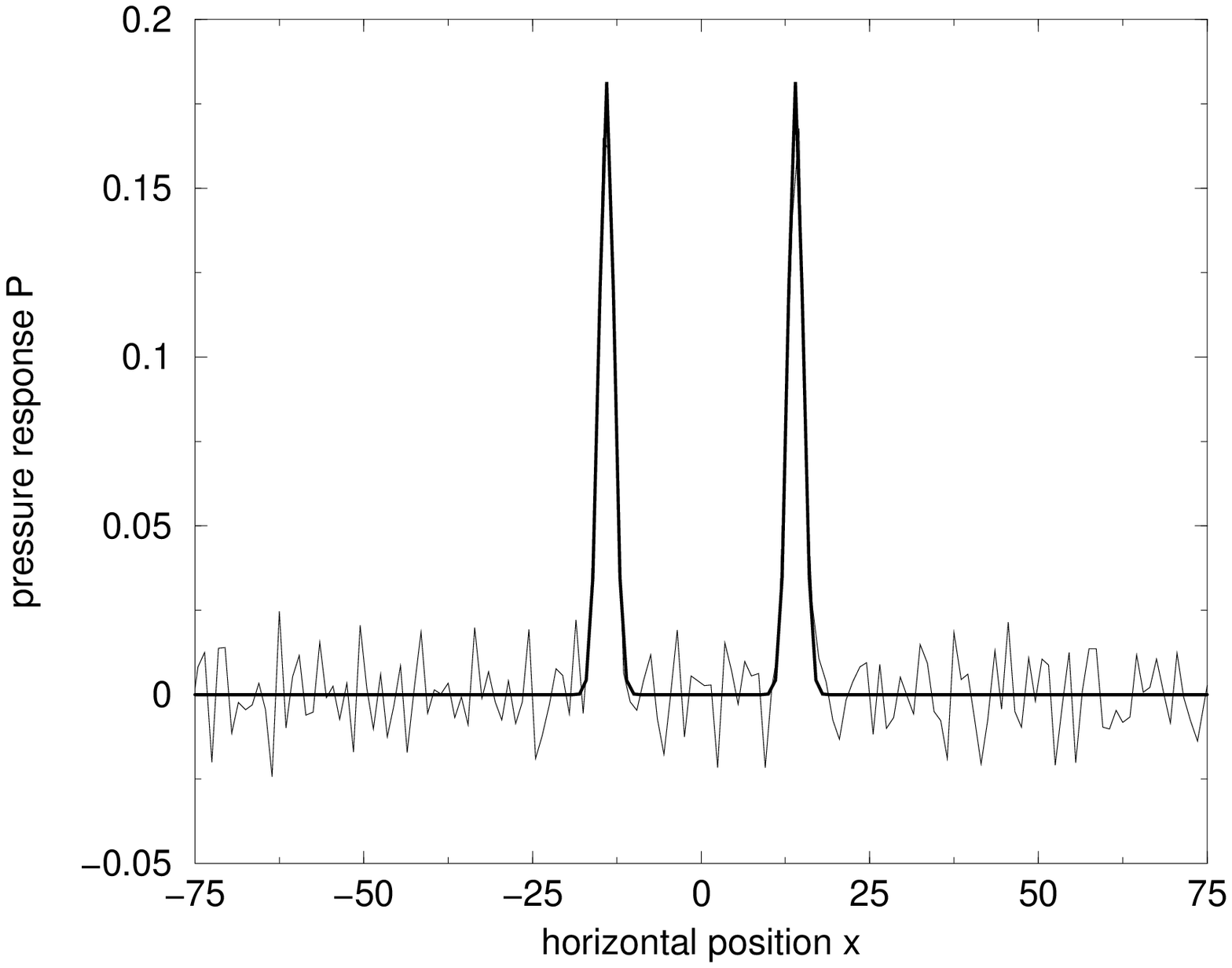}
\hfill
\epsfxsize=0.43\linewidth
\epsfbox{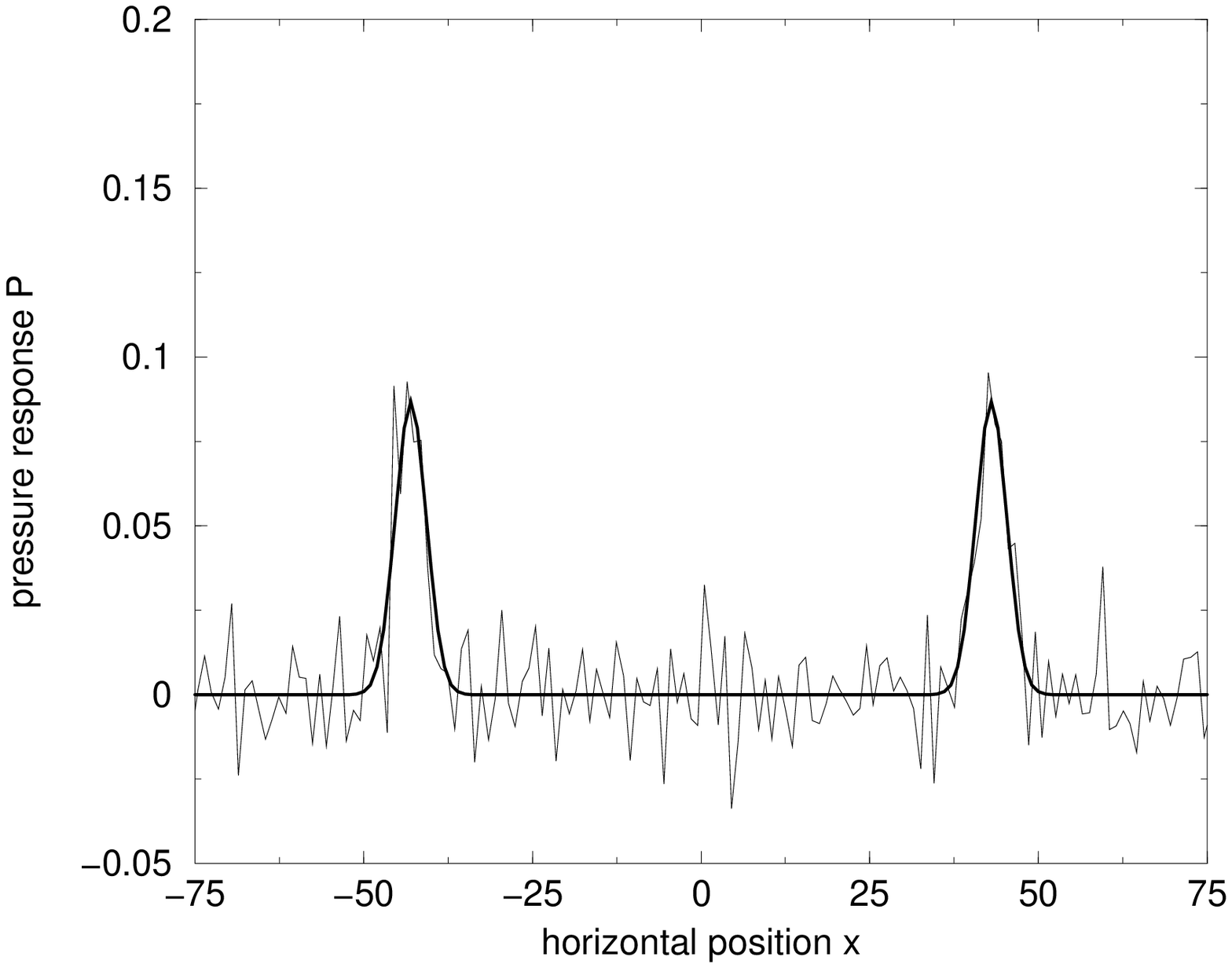}
\end{center}
\caption{
Vertical pressure profile $P(x)$ in response to a unitary force localized at
$x=z=0$, measured at $30$ (left) and $90$ (right) grain layers depth, i.e.
at $z=95.8$ and $z=147.7$. The thin lines are numerical data averaged over
$3461$ different samples. The bold ones are the corresponding gaussian fits
$G(x)$ which give the peak positions $\pm x_p$ and width $W$. As the depth
$z$ increases, the peaks move away from each other, and get smaller and wider.
These data were computed with $\mu=0.3$.
}
\label{rep30et90}
\end{figure}

In this letter we would like to present features of the stress response
function of this model. Gravity is switched off. All grains of the top surface
$z=0$ are overloaded with a unitary vertical force $f_z=1$, except for the
central bead at $x=0$ on which we apply $f_z=2$, see figure \ref{reseau}
(right). For a given value of the
friction coefficient, all contact forces are computed and averaged over
typically few thousands of samples. The uniform unitary confining overload is
subtracted. We call $P(x,z)$ the resulting vertical pressure at point $x$
and depth $z$. Note that this is not the standard response procedure, which
should have been the following: (i) apply a uniform overload at the top and
solve for all contact forces in the layer; (ii) add a small extra force at
$x=0$, and solve for the new forces, keeping the \emph{same} random numbers
for each grain; (iii)~subtract the two previous stress profiles to get the
response. Of course, this would have been much slower. To justify our
alternative procedure, we checked that, as long as $\mu$ is not too large,
the percentage of grains that have to change their random number in step (ii)
in order to satisfy the friction conditions with the new forces is reasonably
small -- of the order of $0.5\%$ for $\mu=0.1$, $2.4\%$ for $\mu=0.3$. Besides,
these rearrangements are rather localized and concern preferentially forces
that are small compared to the additional overload.

Two pressure profiles are shown on figure \ref{rep30et90}. They have been
computed on $500\times200$ systems with $\mu=0.3$, and measured at $30$
(left) and $90$ (right) grain layers depth. They have a double peaked
structure which can be well fitted by a double symmetrical gaussian profile
\begin{equation}
\label{fitgaussien}
G(x) = \frac{1}{2\sqrt{2\pi W^2}}
\left [ e^{-\frac{(x+x_p)^2}{2W^2}} + e^{-\frac{(x-x_p)^2}{2W^2}} \right ],
\end{equation}
where $\pm x_p$ are the positions of the peaks and $W$ their widths. As the
depth $z$ increases, the peaks move away from each other, and they also get
smaller and wider. Although residual fluctuations are large -- they regress
like $1/\sqrt{N_r}$, where $N_r$ is the number of realizations -- we are
able to extract a response whose amplitude decreases like $1/\sqrt{z}$.

\begin{figure}[t]
\begin{center}
\epsfxsize=0.43\linewidth
\epsfbox{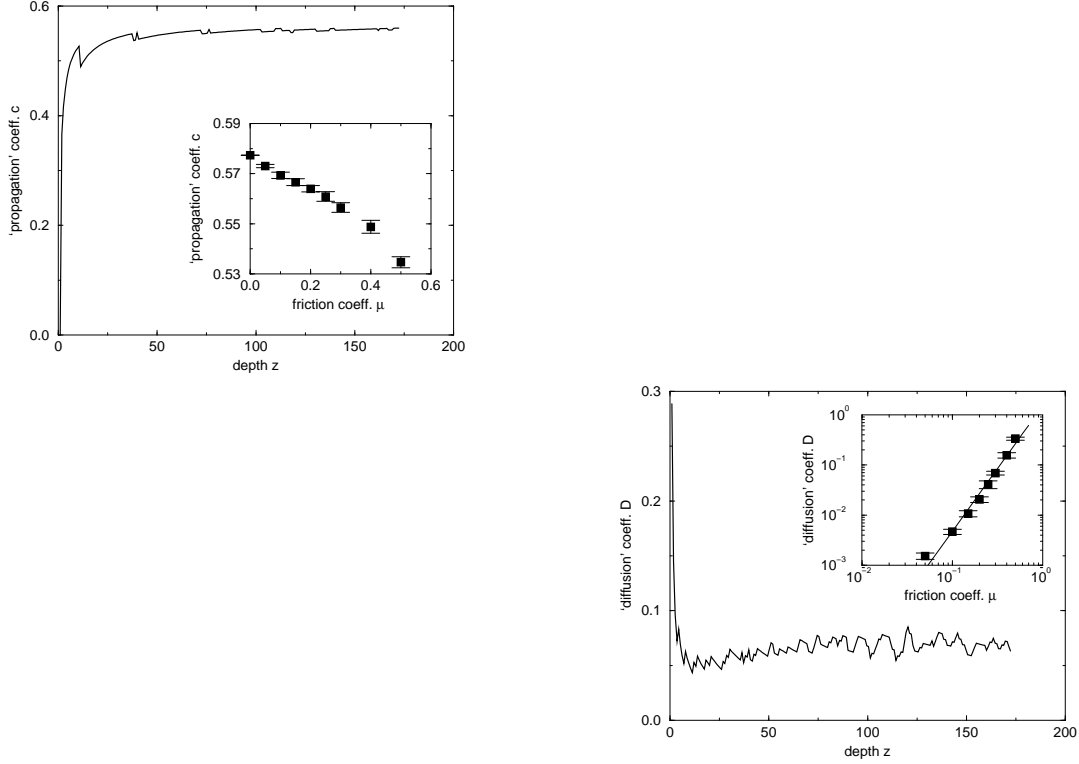}
\hfill
\epsfxsize=0.43\linewidth
\epsfbox{Ddezetdemu.mu=0.3.eps2}
\end{center}
\caption{
\underline{Left}: At large depth, the positions of the response peaks scale
linearly with $z$: $x_p = \pm cz$ where $c$ is the `propagation' coefficient.
This quantity decreases linearly with the friction coefficient $\mu$.
\underline{Right}: Similarly, the peak width grows like in a `diffusion'
process: $W=\sqrt{Dz}$ at large depth. $D$ varies like $\mu^\beta$, with
$\beta \sim 2.5$ (solid line) -- see also figure \protect\ref{cdeD}. The
curve $c(z)$ and $D(z)$ have been computed with $\mu=0.3$.
}
\label{cetD}
\end{figure}

We have studied the evolution of the two parameters $x_p$ and $W$ of the
pressure profiles as a function of depth. As evidenced on figure \ref{cetD},
at large $z$ the ratios $x_p/z$ and $W^2/z$ saturate to some asymptotic values.
In other words, we can define, in analogy to wave propagation and diffusion a
coefficient of `propagation' $c$ and `diffusion' $D$ such that $x_p=cz$ and
$W=\sqrt{Dz}$. We emphasize that these scalings are not compatible with a
homogeneous, anisotropic elasticity analysis. The coefficients $c$ and $D$
depend on the value of the friction $\mu$. As a trivial example,
$c=c_0=\tan30^o$ and $D=0$ at zero friction. $c$ and $D$ as functions of $\mu$
are plotted in the insets of the graphs on figure \ref{cetD}. As one can
expect, $D$ increases with $\mu$. Less intuitively, $c$ is weakly reduced by
the friction. This last result is in agreement with the experimental
observation \cite{manip2D}. More quantitatively, we get the following scalings:
$c_0 - c \propto \mu$ and $D \propto \mu^\beta$, with $\beta \sim 2.5$.

\begin{figure}[t]
\begin{center}
\epsfxsize=0.43\linewidth
\epsfbox{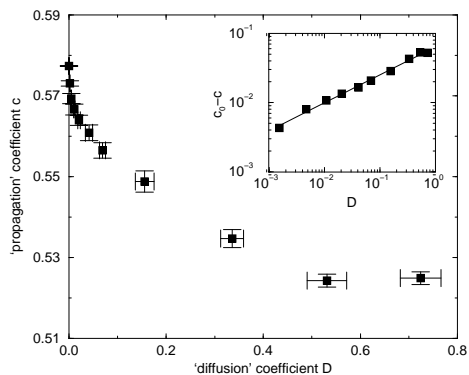}
\end{center}
\caption{This plot shows the `propagation' coefficient $c$ as a function
of the `diffusion' one $D$. Each point corresponds to a different value of
$\mu$. As emphasized by the inset, $c_0 - c \propto D^\alpha$ at small
$D$, were the best fit gives $\alpha = 1/\beta \sim 0.4 \pm 0.01$ (solid line).}
\label{cdeD}
\end{figure}

The features of the pressure response profiles obtained in these calculations
resemble very much those which were predicted in \cite{noisybcc}. The goal of
that paper was to study the role of the disorder on \osl\ equations, i.e. to
close the equilibrium equations by the relation
$\sxx=\eta \left [ 1 + v(x,z) \right ] \szz$, which represents the local
heterogeneities of the granular packing. $v$ is a random noise. Its mean
value is zero and its correlation function is chosen to have the factorable
form $\la v(x,z)v(x',z') \ra = \De^2 g_x(x-x')g_z(z-z')$, where the functions
$g_x$ and $g_z$ are taken short ranged. $\De$ is the amplitude of the disorder. 
When the disorder vanishes, the pressure response function is simply that of a
wave-like equation, i.e. the sum of two $\delta$-functions centered at
$x_p=\pm c_0z$, where $c_0=\sqrt{\eta}$. For finite disorder, although the
exact shape of the response profile is more complicated than a double gaussian
like (\ref{fitgaussien}), it can also be characterized by the position of the
peaks $x=\pm cz$ and the amplitude of their width $W=\sqrt{Dz}$. The
calculation, carried out in the limit of small $\De$, gave
$c_0^2-c^2 \propto \De^2$ and $D-D_0 \propto \De^2$. $D_0$ is the diffusion
coefficient due to the underlying lattice on which the model is defined -- in
the case of the triangular lattice used here, $D_0=0$. In other words, the
quantitative role of disorder that was predicted is the same as what is
obtained here with this numerical model, but the scaling seems to be different.
The theory in \cite{noisybcc} predicts a linear relationship between $c$ and
$D$, at least at small $D$. Figure \ref{cdeD} shows the corresponding plot
from the numerics. As evidenced in the inset we rather get
$c_0 - c \propto D^\alpha$, with $\alpha = 1/\beta \sim 0.4$.

How can we account for this difference? It is actually not easy to identify
the clear correspondence between the random noise $v$ of the theory in
\cite{noisybcc} and the disorder generated by a random choice of a contact
force under Coulombic friction conditions. An important difference, however,
lies in the fact that, in contrast to the noise $v$, the random variable
implemented in these numerics certainly leads to a noise with a \emph{finite
mean value}. This bias is probably dominant in the behavior of $c$, but not
on that of $D$.

To check this point, we plan to use one of the `microscopic justifications' of
the stochastic \osl\ relation proposed in \cite{noisybcc}, which was called the
`three leg model'. In this model, the beads lie on a rectangular lattice and
transmit their forces to their three lower neighbors, the central force being
random. Thus, although no friction condition was explicitly written, the
spirit of the two modellings is very close. The advantage is that we know how
to relate the mean of the random number to the mean propagation coefficient
$\bar{c_0}$, which should allow us to extract the additional effect due to
a zero mean noise. In the three leg model however, no backtracking calculation
was implemented, and eventually negative forces (tractions) appeared at some
finite depth, making the calculation invalid beyond this point. In the future,
we plan to use the highly efficient multi-agent technique at work in the
\textsc{GranuSolve} algorithm that we devised, to test the three leg
model on large scale simulations with the self-consistent condition of
positive forces.

Finally, as mentioned in the introduction of this letter, experiments showed
that strongly disordered frictional granular systems actually have single
peaked response functions \cite{manip3D,repelas,manip2D,Chicago}. In these
simulations, even at very high friction coefficient (up to $1.7$), we were not
able to observe any peak merging. A possible explanation may be that some
additional geometrical disorder is required. We then plan to extend our work
to packings where some random links between grains are opened and thus cannot
transmit any force.

\acknowledgments

This work has been partially supported by a grant to the ''D\'ecouverte''
Research Team (a ``Jeune \'equipe CNRS'). We thank R.P. Behringer, 
R. Bouamrane, J.-P. Bouchaud, S. Galam, G. Reydellet and J.E.S. Socolar
for useful discussions and suggestions.


\end{document}